# The Orthogonal Vulnerabilities of Generative AI Watermarks: A Comparative Empirical Benchmark of Spatial and Latent Provenance

**Authors:** Jesse Yu[1], Nicholas Wei[2]

**Institution:** [1]Millburn High School, [2]Williamsville East High School

**Email:** [1]jesse.gift.yu@gmail.com, [2]nicholas.m.wei3@gmail.com

## Abstract

As open-weights generative AI rapidly proliferates, the ability to synthesize hyper-realistic media has introduced profound challenges to digital trust. Automated disinformation and AI-generated imagery have made robust digital provenance a critical cybersecurity imperative. Currently, state-of-the-art invisible watermarks operate within one of two primary mathematical manifolds: the spatial domain (post-generation pixel embedding) or the latent domain (pre-generation frequency embedding). While existing literature frequently evaluates these models against isolated, classical distortions, there is a critical lack of rigorous, comparative benchmarking against modern generative AI editing tools. In this study, we empirically evaluate two leading representative paradigms, RivaGAN (Spatial) and Tree-Ring (Latent), utilizing an automated Attack Simulation Engine across 30 intensity intervals of geometric and generative perturbations. We formalize an "Adversarial Evasion Region" (AER) framework to measure cryptographic degradation against semantic visual retention (OpenCLIP > 75.0). Our statistical analysis ($n = 100$ per interval, $MOE = \pm 3.92\%$) reveals that these domains possess mutually exclusive, mathematically orthogonal vulnerabilities. Spatial watermarks experience severe cryptographic degradation under algorithmic pixel-rewriting (exhibiting a 67.47% AER evasion rate under Img2Img translation), whereas latent watermarks exhibit profound fragility against geometric misalignment (yielding a 43.20% AER evasion rate under static cropping). By proving that single-domain watermarking is fundamentally insufficient against modern adversarial toolsets, this research exposes a systemic vulnerability in current digital provenance standards and establishes the foundational exigence for future multi-domain cryptographic architectures.

## 1. Introduction: The Imperative for Digital Provenance

We are currently witnessing the rapid, largely unregulated spread of open-weights generative AI. While this technology unlocks immense creative potential, it simultaneously destabilizes mechanisms of digital verification. Automated disinformation, AI-generated images submitted as evidence in legal proceedings, and non-consensual synthetic media are rapidly evolving into systemic real-world vulnerabilities. In response, technology consortia and AI developers have begun deploying invisible watermarks to embed verifiable proof-of-origin into synthetic media. However, robust digital provenance, the ability to accurately track and verify an image's origin through an adversarial laundering pipeline, remains an unsolved computational challenge.

To establish this provenance, a cryptographic payload must survive the journey from the point of generation to the final observer. Modern invisible watermarking algorithms operate strictly within one of two distinct mathematical manifolds:

- **Spatial (Pixel-Based) Watermarks:** Post-generation deep learning models that embed high-frequency binary payloads directly into the visible color pixels of a synthesized image.
- **Latent (Generative) Watermarks:** Pre-generation models that bake continuous cryptographic signatures directly into the initial Fourier frequencies (the underlying wave-like "blueprint") of the AI's starting noise, long before the final image is synthesized.

A critical empirical gap exists within current literature: these methods are predominantly evaluated against basic, isolated classical attacks (e.g., standard cropping or brightness adjustments). Real-world adversaries utilize generative AI editing tools to systematically alter the mathematical structure of an image while preserving its visual semantics. This paper systematically benchmarks a leading spatial model against a leading latent model to formally map their respective boundaries of failure against both classical geometric constraints and modern generative AI editing suites.

## 2. Related Work

The pursuit of digital provenance has evolved significantly alongside the rise of Generative Adversarial Networks (GANs) and Diffusion models, bifurcating into two distinct embedding philosophies.

### 2.1 The Spatial Paradigm and Deep Learning Steganography

Traditional digital steganography relied on classical signal processing, such as Least Significant Bit (LSB) manipulation. The advent of deep learning introduced highly robust spatial architectures which utilize adversarial training to embed dense payloads into the spatial layout of an image. Current open-source spatial SOTA models, such as RivaGAN, utilize attention-based encoders to route payloads into high-variance pixel textures. Recent closed-source industrial advancements, most notably Google DeepMind's SynthID, also operate heavily within pixel-level perceptual optimizations. These spatial models theoretically excel at surviving classical signal distortions because the payload is redundantly distributed across the 2D pixel grid. However, recent theoretical studies suggest spatial marks struggle against deep-learning-based algorithmic attacks, as diffusion models naturally act as low-pass filters that overwrite localized high-frequency pixels.

### 2.2 The Latent Paradigm and Diffusion-Based Provenance

To counter the vulnerability of pixel-level editing, researchers introduced Latent watermarking. Models like *Tree-Ring* shift the cryptographic payload entirely out of the pixel space and into the mathematical foundation of the AI generator. Tree-Ring embeds a cryptographic key into the continuous Fourier frequency domain of the initial diffusion noise vector. Because the watermark dictates the global semantic layout of the final image, it demonstrates profound theoretical resilience against localized generative alterations. However, because Fourier transforms rely on strict global geometry, these models are hypothesized to be inherently vulnerable to physical grid misalignment.

### 2.3 Industrial Advancements and the Benchmarking Gap

Recent industrial efforts, such as Google's SynthID and the Coalition for Content Provenance and Authenticity (C2PA) metadata standards, represent major strides in commercial watermarking. However, because proprietary models like SynthID are largely closed-source, independent comparative benchmarking remains difficult. Consequently, open-source representatives like Tree-Ring and RivaGAN serve as critical proxies for evaluating the foundational vulnerabilities of latent and spatial manifolds. Few studies have conducted a direct, large-scale comparative analysis across both generative and geometric attacks using a standardized visual-utility framework.

## 3. Methodology and Experimental Design

### 3.1 Data Source and Generation Baseline

To ensure unbiased and standardized textual representations, we sampled unique text prompts uniformly from the *DiffusionDB* dataset (specifically the `2M-first-1k` split). We generated a base set of 4,000 images: 2,000 images natively embedded with Tree-Ring latent Fourier signatures during generation, and 2,000 clean images subsequently watermarked post-generation using pre-trained RivaGAN weights.

*Model Selection Justification:* Stable Diffusion v1.5 was utilized as our base generative model. While newer architectures (e.g., SDXL, SD3, and Flux) are prevalent in 2026, SD v1.5 remains the foundational architecture for the vast majority of open-source adversarial laundering pipelines due to its low VRAM requirements and extensive ControlNet/LoRA ecosystem. Establishing a baseline on v1.5 ensures our attack simulations accurately reflect the accessible tools utilized by real-world threat actors.

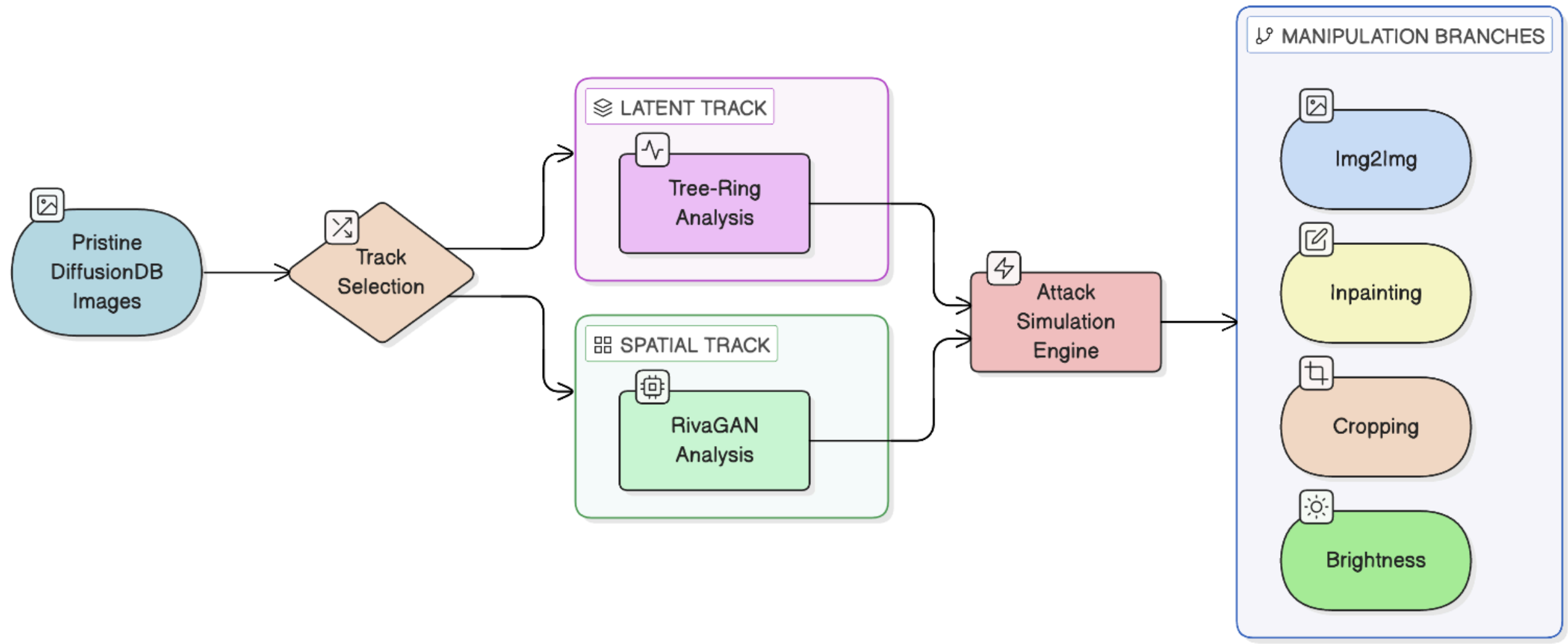

*Fig 1: The Automated Attack Simulation Engine*

### 3.2 The Attack Simulation Engine and Hyperparameters

We developed an automated Python pipeline utilizing `diffusers` and `torchvision` libraries to subject these images to isolated attacks across 30 equal intensity intervals ($N = 30$).

- **Image-to-Image (Img2Img) Translation:** Rewriting the global pixel structure via Stable Diffusion resampling. We utilized a `DDIMScheduler` for 15 inference steps, scaling the noise strength linearly from 0.01 to 0.95.
- **Semantic Inpainting:** Masking and replacing localized regions of the image via a generative inpainting pipeline (15 inference steps). To standardize the area ratio removal without introducing the variability of saliency algorithms (e.g., Segment Anything Model), masks were generated as deterministic, centered geometric bounding boxes proportional to the target area ratio (5% to 60%).
- **Geometric Cropping:** Utilizing `torchvision.transforms` and standard PIL operations to center-crop outer boundaries, destroying spatial grid alignments (removing 5% to 90% of the area).
- **Brightness Adjustments:** Standard pixel intensity scaling utilizing `ImageEnhance.Brightness` (factors of 1.0 to 3.0).

### 3.3 The Adversarial Evasion Region (AER) Framework

An attack is only considered computationally "successful" if it scrubs the watermark while preserving the visual utility of the image. Due to architectural differences, provenance survival is normalized to a $0.0 - 1.0$ scale via distinct extractions:

- **Tree-Ring (Latent):** Measured using Mean Squared Error (MSE) against the original cryptographic seed: $\max\left(0, 1.0 - \frac{MSE}{\sigma^2}\right)$, where $\sigma^2$ is the empirical pixel variance (observed at $\approx 5000$).
- **RivaGAN (Spatial):** Measured by the byte-accuracy of the binary payload.
- **Fidelity Metric:** We measured semantic retention using the OpenCLIP (ViT-B-32) cosine similarity score (scaled 0-100) to evaluate high-level visual concepts.

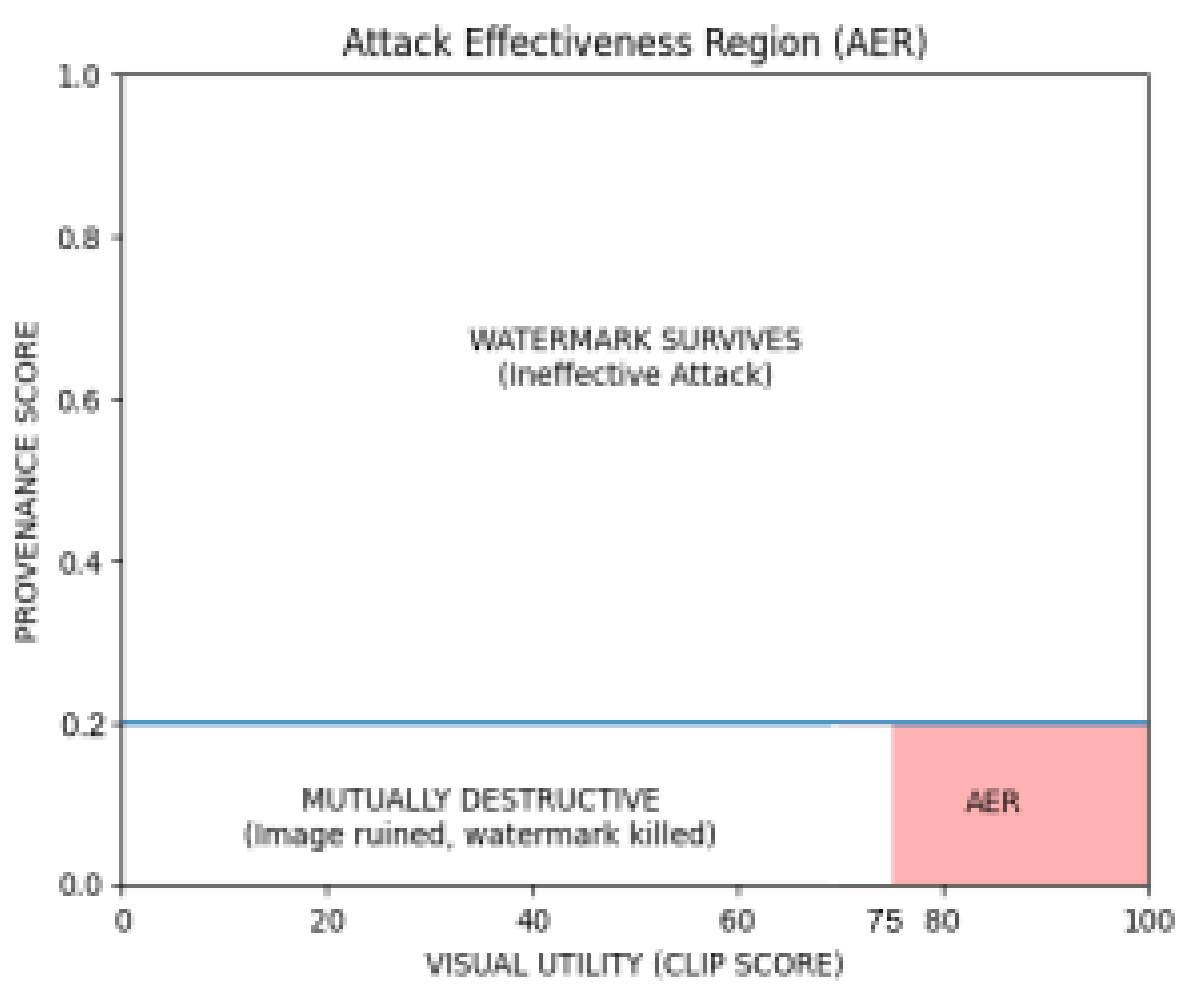

We formalize a successful evasion as a state occurring within the **Adversarial Evasion Region (AER)**, a state where the attack reduces the Provenance Score below a **Critical Evasion Threshold (CET) of 0.20**, while maintaining high visual utility (CLIP similarity > **75.0**).

*Fig 2: AER Representation*

### 3.4 Statistical Bounds and Uncertainty

For each attack intensity interval, we tested an independent sample of random images ($n = 100$). Utilizing the Central Limit Theorem ($n \geq 30$), and an estimated population standard deviation of 0.20, our Margin of Error (MOE) at a 95% confidence interval is mathematically bounded at:

$$MOE = 1.96 \times \left(\frac{0.20}{\sqrt{100}}\right) \approx \pm 3.92\%$$

Furthermore, because DDIM inversion is an approximate ODE solver rather than a perfectly lossless transformation, unattacked latent images naturally yield a baseline provenance noise floor of $\sim 0.95$ rather than a perfect 1.0. This natural generative variance is accounted for within our statistical bounds.

## 4. Results: The Discovery of Orthogonal Vulnerabilities

The empirical benchmarking across thousands of single-vector permutations revealed a stark divergence in the failure modes of the spatial and latent paradigms. Rather than sharing weaknesses, their vulnerabilities are mathematically **orthogonal** (mutually exclusive).

| Model | Attack | AER Rate (%) |
| --- | --- | --- |
| **RivaGAN** | Control | 0.00 |
| **RivaGAN** | Brightness | 22.00 |
| **RivaGAN** | Crop | 22.67 |
| **RivaGAN** | Img2Img | 67.47 |
| **RivaGAN** | Inpainting | 66.80 |
| **Tree-Ring** | Control | 0.00 |
| **Tree-Ring** | Brightness | 0.00 |
| **Tree-Ring** | Crop | 43.20 |
| **Tree-Ring** | Img2Img | 17.73 |
| **Tree-Ring** | Inpainting | 10.27 |

*Fig 3: AER Evasion Rates*

### 4.1 Spatial Fragility: Generative Overwriting

The spatial paradigm (RivaGAN) demonstrated relative stability against non-destructive physical changes, suffering a moderate 22.67% AER against basic Cropping.

However, RivaGAN exhibited severe cryptographic degradation when subjected to algorithmic generative attacks. Under Img2Img translation, RivaGAN experienced a **67.47% Adversarial Evasion Rate**, and a **66.80% AER** under generative Inpainting. Because spatial algorithms embed high-frequency pseudo-random noise broadly across the pixel space, the Stable Diffusion adversarial attack effectively treats the spatial payload as generic image static. The generative model systematically "denoises" and overwrites the pixel-level watermark while perfectly preserving the underlying semantic image structure.

### 4.2 Latent Fragility: Geometric Desynchronization

Conversely, the latent paradigm (Tree-Ring) proved highly robust against the exact algorithmic attacks that degraded the spatial model. Tree-Ring suffered only a 17.73% AER under Img2Img attacks and a negligible 10.27% AER under Inpainting. Because the latent signature is baked into the foundational Fourier blueprint of the image, semantic pixel-rewriting fails to dislodge the core frequency signature.

However, Tree-Ring suffered systemic geometric failure. Under standard static cropping, Tree-Ring's adversarial evasion rate spiked to **43.20%** (nearly double that of the spatial paradigm). The latent extraction process relies on Fast Fourier Transforms (FFTs), which are heavily dependent on global 2D grid alignments. The moment the outer edges of the image are cropped, the spatial grid shifts, causing the continuous frequency rings to mathematically desynchronize during extraction.

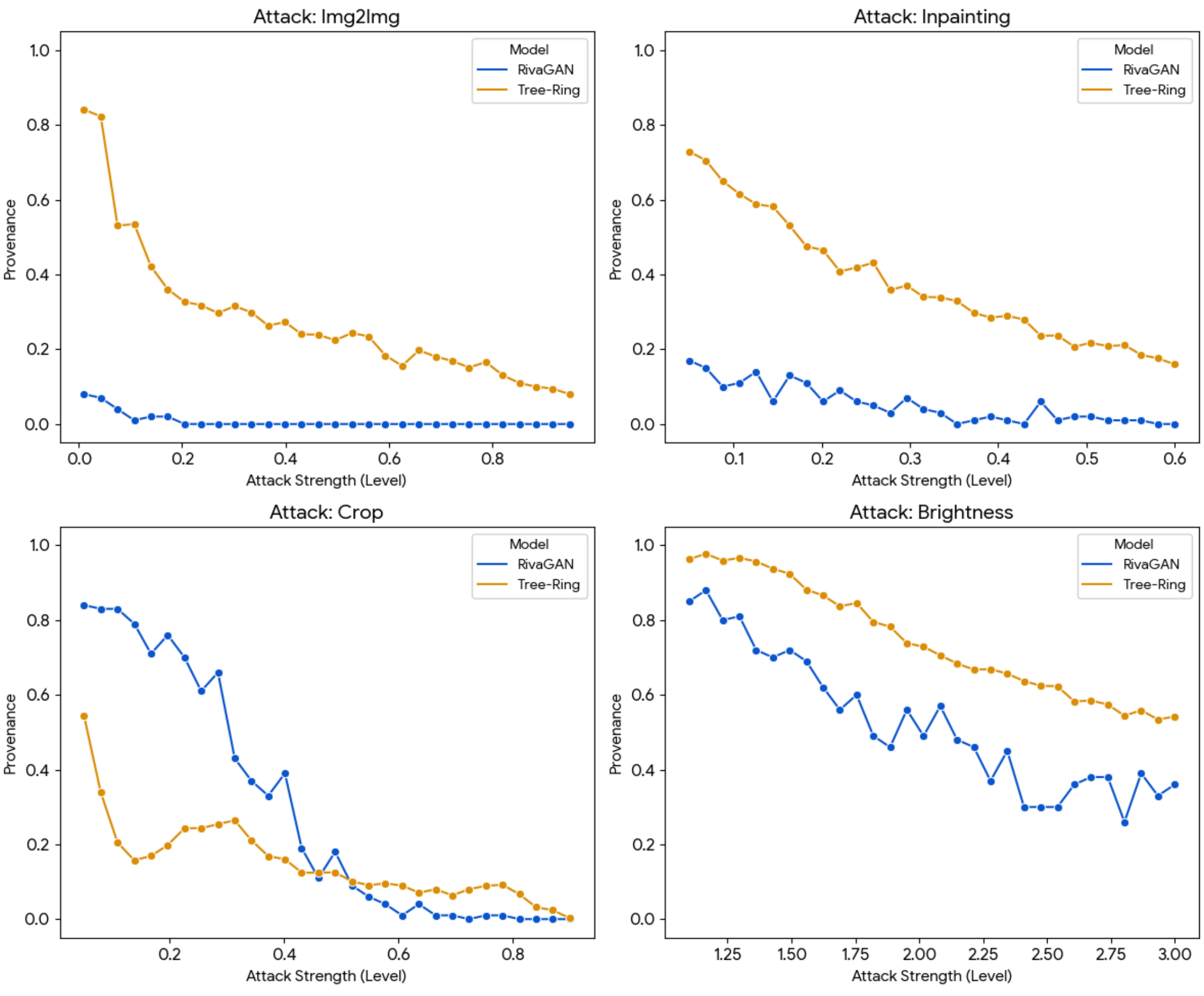

*Fig 4: Comparative Degradation Line Graph*

## 5. Limitations

To maintain scientific rigor, we formally acknowledge several constraints within our methodology:

**Model Specificity vs. Generalization:** This study benchmarked RivaGAN and Tree-Ring as the preeminent open-source representatives of the spatial and latent paradigms. While the underlying math strongly suggests these vulnerabilities are inherent to the embedding manifolds, further research is required to confirm if these orthogonal failure rates apply uniformly to all spatial and latent architectures.

**Computational Constraints and Base Model Scaling:** Evaluating images across a highly granular attack matrix requires significant GPU inference time. Consequently, our simulations were conducted on Stable Diffusion v1.5. Evaluating newer, highly parameterized models (e.g., SD3 or Flux) may yield different noise-filtering behaviors during Img2Img laundering, which warrants future exploration.

**Attack Masking Nuance:** Our inpainting attacks utilized deterministic geometric bounding boxes. Real-world adversaries may utilize semantic segmentation (e.g., the Segment Anything Model) to target specific subjects in an image, which could impact global Fourier frequencies differently than geometric excision.

**Epistemic Uncertainty:** Tree-Ring detection relies on DDIM inversion, which is not perfectly lossless due to approximation drift in the reverse ODE steps. Unattacked latent images naturally yield a baseline provenance score of $\sim 0.95$ rather than a perfect 1.0, though this variance is accounted for within our statistical bounds.

## 6. Conclusion and Future Works

As demonstrated by our rigorous iso-utility benchmarking pipeline, evaluating generative watermarks against a narrow set of classical distortions projects a dangerously incomplete picture of real-world security. Fundamentally, single-domain provenance systems possess inherent, inescapable structural blind spots. We formally conclude that:

- **Spatial watermarks** are inherently fragile to generative AI overwriting (Img2Img and Inpainting).
- **Latent watermarks** are inherently fragile to grid-shifting and geometric misalignment (Cropping).

**Future Works: The Exigence for Dual-Layer Synergy**

The orthogonal nature of these vulnerabilities presents a profound theoretical opportunity for future cybersecurity research. Because Latent marks survive what Spatial marks fail, and Spatial marks survive what Latent marks fail, the industry must explore **multi-domain, dual-layer combination watermarks**.

If researchers can intelligently stack a pixel-based payload directly into the non-interfering null spaces of a latent frequency blueprint, the resulting architecture could theoretically survive both geometric and generative attacks. However, we hypothesize that naively combining these methods will result in structural signal interference. Future studies must engineer intelligent, adaptive routing algorithms to resolve this internal signal collision.

Furthermore, because real-world adversaries utilize compounding tools to launder images, future research must shift focus away from single-vector attacks and begin evaluating watermarks against complex, combinatorial multi-vector threat models (e.g., sequentially chaining Inpainting, Img2Img, and Cropping). Single-domain watermarking is no longer sufficient; the field must move toward synergistic defenses.